\documentstyle [12pt,psfig]{article}
\begin{document}
\title{Luminosity Distributions within Rich Clusters - I: \\
A Ubiquitous Dwarf-Rich Luminosity Function ?}
\author{Rodney M. Smith$^{1}$, Simon P. Driver$^{2}$ and Steven Phillipps$^{3}$\\
$^{1}$Department of Physics and Astronomy, University of Wales \\
College of Cardiff, PO Box 913, Cardiff CF2 3YB\\
$^{2}$Department of Astrophysics, School of Physics, \\
University of New South Wales, Sydney, NSW 2052, Australia\\
$^{3}$Astrophysics Group, Department of Physics, University of Bristol,\\ 
Tyndall Avenue, Bristol BS8 1TL
}

\maketitle

\begin{abstract}
From deep CCD observations of the cluster Abell 2554 we have recovered the
cluster's luminosity distribution over a wide range of magnitude
($-24 < M_{R} < -16$). We compare the derived A~2554 cluster luminosity
function (at 
redshift 0.1) with that of the local Coma Cluster (A~1656)
and the more distant 
($z = 0.2$) cluster A~963. The distribution is remarkably similar for these 
three clusters of comparable richness and morphology. 
All show a flat ($\alpha = -1.0$) 
luminosity function for the giant galaxies ($-24 < M_{R} < -19.5$) which 
exhibits a sharp upturn ($\alpha \approx -1.7$) at some 
intermediate magnitude ($M_{R} \simeq -19$) and continues to rise to the
limits of existing data. We suggest that such a luminosity function may be ubiquitous among rich 
clusters and that a similar form may apply for poorer clusters and
possibly the field as well. The three cluster
dwarf LFs are seen over a 
range of lookback times covering a quarter of the age of the universe. 
Therefore the similarity between the three measured LFs seems to rule out
strong evolution of the dwarf populations in rich cluster environments,
at least out to z = 0.2, unless richness effects conspire to conceal
evolutionary changes.

\end{abstract}

\section{Introduction}

There has recently been a renewal of interest in the galaxy luminosity function
(LF), in particular in its faint end. This has been driven partly by improved
data (ie. we can now observe fainter galaxies; see eg., Bernstein {\it et al.}
1995),
partly by increased interest in dwarf galaxies and their number density {\it
per se} (eg., Ferguson \& Binggeli 1995), and partly by the recognition that
low luminosity galaxies and their evolution almost certainly play a vital
cosmological role in, for example, explaining the large numbers of galaxies
counted at faint magnitudes (eg., Phillipps \& Driver 1995; Ellis {\it et al.} 
1996).

Although in the most general context one would like to know the luminosity
function of field galaxies to very faint levels, this is observationally
limited, primarily by the requirement for individual redshifts and by the
unavoidable predominance of brighter galaxies in magnitude limited samples (see
Driver \& Phillipps 1996). One may also worry about the possible selection
effects in the faint end of the local field LF (McGaugh 1994; Schade \&
Ferguson 1994; Phillipps \& Driver 1995). One is therefore led to
determinations of the LF of rich clusters of galaxies and several groups
have recently presented cluster LFs down to absolute magnitudes previously
sampled only in our immediate neighbourhood (see Thompson \& Gregory 1993;
Driver {\it et al.} 1994b; Bernstein {\it et al.} 1995; De Propris {\it et al.}
1995). Here we add to these recent results via a 
deep CCD imaging survey of the cluster Abell 2554 which allows us to determine 
the LF to over 6 magnitudes below the characteristic magnitude of giant 
galaxies (ie. $L = 0.003 L_{*}$).

We also present a first attempt to  examine the evolution of the rich
cluster LF at faint levels. Very extensive redshift surveys carried out in the
last few years (Ellis {\it et al.} 1996; Lilly {\it et al.} 1995) have supplied evidence for
a change in the field galaxy LF - either steepening, shifting or both -
for moderately low luminosity dwarfs (see also Broadhurst, Ellis \& Shanks
1988; Colless 1995). Evidence - one way or the other - for evolution of cluster
dwarfs is still lacking, however. The well-known Butcher-Oemler (1984) effect
is primarily a spectroscopic or morphological evolution (eg., Couch \& Sharples
1987) whose effect on the LF has not, to our knowledge, been investigated
(though see Barger {\it et al.} 1996 for a related discussion). In
any case, the known cases are restricted to fairly luminous galaxies.
Here we discuss our new observations of A~2554 at redshift
$z = 0.1$ and compare these with similar observations of A~963 ($z = 0.2$)
(from Driver {\it et al.} 1994b) 
and published results on the Coma cluster at $z =
0.02$ (eg. Godwin \& Peach 1977, Godwin, Metcalfe \& Peach 1983). 
In a model with $\Omega = 1$, $H_{0} = 50$km~s$^{-1}$Mpc$^{-1}$ (ie. $h = 0.5$
in the usual notation), as assumed
herein for consistency with previous work,
this gives us a range in look back times of a quarter of the age of
the universe or $\simeq 3.3$ Gyr.

In Section 2 we describe the observations and data analysis whilst in 
Section 3 we detail the recovery of the LF of A2554. Section 4 describes our 
analysis of the structure of A2554 and in Section 5 how this cluster
compares with other clusters whose LFs are known. Section 6 contains 
a brief description of the relevance of these observations to the
evolutionary scenarios explaining dwarf galaxy evolution and the faint blue
galaxy excess seen in the number counts.

\section{The Data}

\subsection{Observations}

A~2554 ($23^{h}~09^{m}~48^{s}, -21^{o}~45'$) is a richness class 3 cluster of
Bautz-Morgan type II: at $z = 0.106$ (Abell 1958; Leir \& van den Berg 1975;
Abell, Corwin \& Olowin 1989). The observations were taken by RMS in 1994
July at the Anglo-Australian Telescope, as part of a larger programme of
cluster photometry (see Phillipps, Driver \& Smith 1995a,b). 
Data were obtained using the
f/1 focal reducing optics and $1024 \times 1024$ pixel Thomson CCD (see eg.,
Turner {\it et al.} 1993 for details of this system). The pixel size is $0.98''$
giving a total field of view $17'$ square. Note that this corresponds to almost
exactly 3 Mpc at the cluster distance (600 Mpc for our choice of cosmological
parameters). A grey scale representation of our image of A~2554 is shown in
figure 1. The seeing for this exposure was 2''.

Multiple short exposures (75 or 100 seconds each) were obtained to aid defect 
removal
and flat fielding.
Frames were `jittered' by 10 arcseconds between exposures
to minimize systematic effects on the pixel scale.
The total exposure time was 4850 seconds in the Kron-Cousins
$R$ band. After standard bias subtraction, a master flat field was constructed
from a median stack of all the disregistered cluster frames (both for this
cluster and for others observed in the same run), as in Driver {\it et al.}
(1994a,b). After flattening by this master flat, the individual frames of the
cluster area were registered and co-added. This reduces the pixel-to-pixel sky
noise to near the level of the Poisson counting noise (0.5 \%), but may
leave residual larger scale variations in the background level which can be
important for faint object detection (see eg., Davies {\it et al.} 1994). These
variations were removed by subtracting a spatially median filtered version of
the data (with box size 64 pixels), as discussed in Schwartzenberg {\it et al.}
(1995). Region to region variations (say on arc minute scales) are then reduced
to 0.2 \% or less. Blank (ie. off cluster) frames, including
in particular ones in the well studied South Galactic Pole region (eg., Couch,
Jurcevic \& Boyle 1993), were also taken, during the same night and similar
seeing, with
an identical set-up as that for A 2554. The SGP field was reduced 
in exactly the same way as the cluster frames. 
Exposures of Landolt (1993)
standard stars throughout the night showed that conditions
were photometric to about 0.02 magnitudes. Conversion to the standard
$R_c$ system was also carried out using the stellar observations.
\\

\subsection{Image Detection}

The resulting data frame was processed using 
the PISA image detection package in the STARLINK suite of astronomical
data reduction software. This package utilizes a connected pixel algorithm 
to detect objects containing more than a limiting number of pixels above
a user-defined threshold. Overlapping objects were split into their 
constituent objects using a deblending algorithm. To prevent this effect
occurring in the haloes of bright stars, such regions, with a generous
allowance for the outer low-level wings of the stellar profile, were
excluded from the analysis. For this survey an isophotal magnitude was 
calculated using limits for the detection
of at least 4 pixels in area, with
each pixel above a limit of $1\sigma$ above sky (the sky `noise' $\sigma$ 
being determined directly from the histogram of pixel values). For the present
data this threshold corresponds to 26.9 $R\mu$ 
(0.5\% of sky, as noted earlier). 
After object detection, a further (conservative) $7.5\sigma$ cut was applied to eliminate
possible noise `objects' (cf. Driver {\it et al.} 1994b).
This combination will $not$ result in a magnitude limited sample unless all
galaxies of a given apparent magnitude also all have the same surface
brightness, or at least contrive not to extend into undetectable areas of
parameter space (cf. the discussion in Driver {\it et al.} 1994a). In particular,
fainter galaxy images have to be more compact in order to pass the
signal-to-noise test. (Large diffuse images contain more sky noise). We should
always bear this in mind when assessing the results, especially in comparisons
between different data sets. 
This effect will lead
to an underestimate of the steepness of the faint-end of the LF. This is
supported by the work of Schwartzenberg (1996) on dwarf galaxies
in the Virgo cluster who finds a steeper LF than presented here.
Figure 2 shows all the detections ($\geq 4$ pixels) in
the parameter space of isophotal area versus isophotal magnitude. Solid
symbols indicate the detections with S/N $\geq 7.5$. (The limit is simply a 
diagonal
line in this plot). The upper line plotted on the figure is the ultimate
selection limit for an image of a given number of pixels all at the detection 
threshold. As an indication of the intrinsic parameters involved, figure 2
also shows the loci of images (assumed to have perfect exponential profiles)
of particular central surface brightnesses and scale sizes.
(cf. Schwartzenberg {\it et al.} 1995). Note that the calculated parameters are
not corrected for the
effects of seeing or noise (discussed in Turner {\it et al.} 1993).

\section{Recovering A~2554's Luminosity Distribution}

Typically for deep CCD images a uniform correction factor is applied to convert the 
number of actual galaxies detected to the number per square degree. However, 
for cluster sight-lines an additional consideration is the 
obscuration of fainter galaxies by the brighter cluster members. 
In dense cluster environments a significant fraction (5-40\%) 
of the field of view may be occupied by the more extended brighter cluster 
members, hence progressively fainter galaxies are sampled over a smaller field 
of view. Normally for field sight-lines this crowding effect is 
negligible at magnitudes significantly brighter than the confusion limit, due 
to the paucity of bright foreground galaxies. However this is not necessarily
the case when dealing with cluster sight-lines and/or very faint galaxies
and the importance of including 
this correction depends on the cluster's 
richness, redshift and morphology. The simplest way to account for this 
``diminishing area'' effect is to determine the field-of-view correction for 
each individual galaxy, by summing the cumulative isophotal area covered by all
brighter objects. This assumes that in the case where a faint foreground galaxy
overlaps with a more luminous distant galaxy, the faint galaxy's 
luminosity is essentially lost; it will neither be detected itself nor is its
luminosity likely to add significantly to that of the brighter galaxy
(though see Bristow 1996 for a more detailed discussion of the effect of
image overlap on galaxy counts). 

Figure 3 shows the significance of this correction for both our A~2554
cluster and the SGP blank field sight-lines. We see that at $m_{R} = 23$ 
(our completeness limits) the 
field correction is 1.08 (filled triangles, {\it i.e.} 8\% of the available 
field-of-view
is occupied by galaxies with $m_{R} < 23$), while for the cluster the 
correction is 1.19 (filled squares, {\it i.e.} 19\%). This correction is 
clearly significant, but this simple
method of correcting ignores the radial 
distribution of the cluster galaxies. 
Given that the luminous galaxies are 
normally clustered around a central core then it follows that the central 
region is more likely to be obscured than the outer regions. 
Figure 3 therefore illustrates the required correction for the core region 
(short dashed) and that for 
an outer region close to the frame edges (long dashed). At $m_{R}=23$
these are factors of 1.25 and 1.16 respectively (adopting an effective mean 
distance modulus to A2554 of, $D_{A2554} = 39.2$). That the outer
annulus still has a diminishing area correction greater than that of the blank
field, implies that the cluster size is greater than our total field-of-view. 
The open squares shown in Figure 3 represent
the ratio of radially corrected number-counts (using three radial ranges; core, 
surrounding annulus and edges) to non-corrected counts for 
the cluster sight-line.
While this radial correction generally produces marginally
higher counts than the uniform diminishing area correction, some additional 
scatter is introduced because of (a) limited number statistics and (b)
the  distribution of cluster galaxies not being perfectly circular. We 
therefore use the mean cluster area correction in what follows.
The radial profile of A~2554 is discussed later in section 6. 

The final corrected number-counts for the A~2554 field are shown in their usual
form in Figure 4 (upper panel). The SGP field has also been uniformly
corrected for the ``diminishing area''. Overlain, as a solid line on Fig. 4, 
is our best fit to the R-band counts from Metcalfe {\it et al.} (1995)
corrected to $R_{c}$ ($= r_{F}+0.1$), {\it viz.}
$log N = 0.377R_{C}-4.615$. 
The SGP counts, our comparison 
field, lie marginally below those of Metcalfe {\it et al.} at the faintest 
limit. This is to be expected as Metcalfe {\it et al.}'s data reach to fainter
isophotal detection thresholds and are hence less effected by 
surface brightness selection effects (ie. they include slightly lower surface
brightness objects at a given magnitude). The A~2554 data and SGP data have 
essentially identical isophotal depth (and are both complete to $R \sim 22.5$
at least), so no (galaxy population 
dependent) corrections for surface brightness selection effects need to be 
considered.

In order to obtain the cluster-only galaxy counts and hence the LF
we need to subtract the background counts (c.f. Driver {\it et al.} 1994b). 
We can do this in two ways.
Firstly, we can assume that the background counts are well known. This
has the advantage of minimal Poisson noise since we can use a best fit to
the extensive published data, but may have systematic
effects because of differences in detection procedures or from field to
field fluctuations. Second, we can use our own 'blank' field to determine the
background/foreground contamination. This has the obvious advantage of 
identical image detection and processing, but increases the Poisson noise (by 
about $\sqrt{2}$). It will also fail to remove the possibility of field to 
field variations,
unless the line-of-sight is close to the line-of-sight of the cluster. In this
case our reference blank SGP field lies approximately 15 degrees from A~2554.
In addition to the two methods outlined above
we could also use the variations in number density across our field itself to
determine the background level. However for this method we would require a 
larger field-of-view 
(recall from Figure 3 that the outer radial correction for the 
diminishing area in A~2554 is still greater than that for the SGP field
implying that the cluster population extends beyond the edge of our 
field).

Figure 4 (lower panel) shows the application of the first two techniques, based
on the counts of Metcalfe {\it et al.} (1995) and our SGP field
respectively. The two methods imply closely similar
results for the luminosity function of
A~2554, with most of the discrepancy at either the very bright or very
faint ends of the luminosity distribution. Overall the agreement is excellent,
and within the errors of both analyses. Note that the errorbars include the 
assumption of Poissonion statistics in the numbers of detections
added in quadrature to a uniform
10 \% error. The latter allows for the possibility of a 
0.1 magnitude systematic error in our photometric zero points and/or 
field-to-field variations in the background numbers.

The marginal discrepancies at the bright and faint ends can readily be 
explained.
At bright magnitudes the SGP is more susceptible to field-to-field variations
than the more extensive data given in Metcalfe {\it et al.}, 
due to the SGP field's small area.
As noted above,
at our faintest magnitudes the Metcalfe {\it et al.} data are liable 
to include some galaxies with surface brightnesses too low to be included in 
either the A~2554 or SGP data. 
Based on this, we can adopt a `best buy' model which uses
the A~2554 -- Metcalfe
data points for $M_{R} < -19.5$ and the A~2554 -- SGP points for $-19.5 < M_{R}$.
The solid line 
shows a flat Schechter (1976) function with $M_{R}=-22.3$, $\alpha = -1$ 
normalised to match the $M_{R} < -19.5$ data of A~2554 -- Metcalfe.

If we do consider the excess number of galaxies in the central regions compared to the edges, then a rather similar shaped LF is recovered (though, of course, 
the normalization is not the same as we are subtracting off some cluster 
members), up until the point where the small number statistics begin to 
dominate. Apart from confirming the steep dwarf part of the LF, this also
indicates that the radial fall-off for different luminosity
galaxies is not all that dissimilar (ie. little luminosity segregation). 

\section{The Morphological Structure of A~2554}

Figure 5 shows the radial distribution of cluster galaxies 
about the cluster centre, 
divided into two magnitude intervals. These intervals have been 
chosen such that the sample is divided at the point where the luminosity 
distribution shows the distinctive upturn. The radial profile of the more 
luminous galaxies rapidly drops off away from the centre but does not reach the
density of the field within our field-of-view. (In fact the cluster probably
has a more complex underlying structure  
than a simple isothermal sphere, see below). 
The less luminous galaxies also show a gradual fall-off towards the frame
edges but less so than the giants.

We might worry slightly that the dwarf population seems to follow a flatter 
distribution, as this is precisely what one would expect if we had
underestimated the background subtraction due, for example, to a zero
point shift in our cluster field calibration. (Since the cluster galaxy numbers
increase less quickly than the background counts, background errors have a
larger effect at faint magnitudes). In order to examine the 
morphological structure of
the cluster in a little more detail, Figure 6 shows the number-density 
enhancement across the field of view for the giants ($14.0 < R < 20.0$) and 
dwarfs ($20.0 < R < 22.75$) in turn (Figs. 6a and 6b respectively). 

The gray scale plots show the excess number of galaxies over the mean field
value for the bright cluster galaxies (a), bright field galaxies (b),
faint cluster galaxies (c) and faint field galaxies (d). 
The grayscales are
scaled from the mean field value to the mean field value plus 48.0 galaxies 
per $200\times200$ pixel cell. 
The isodensity contour lines are set to:
+6.0, +12.0, +18.0, +24.0, +30.0, +36.0 and +42.0 excess galaxies per 
$200\times200$ pixel cell (Figs 6(a) and (c)) and 
+18.0, +30.0, +42.0, +48.0, +54.0, +60.0 and +66.0
excess galaxies per $200\times200$ pixel cell (Figs 6(b) and (d), scale up by
$\sim 300$ to get to excess galaxies per sq deg). 

Figure 6a therefore shows the 2-D distribution of the cluster giants which is 
seen to be non-symmetrical and to contain significant density structure out to 
the frame edges. For comparison Fig 6b shows the equivalent plot for the SGP 
field where minimal structure is expected (and seen). Fig. 6c shows the dwarf 
distribution which closely follows the distribution of the giants (note the
main central peak and extensions towards the top and right-side).
The equivalent density plot for the fainter field galaxies (Fig 6d.)
shows that both the density and structure of the dwarf distribution 
in A~2554 are indeed significant. We therefore conclude that the dwarf galaxies
follow the distribution for giants, at least approximately, and that background
subtraction errors are not responsible for the large number of dwarfs seen.
The distribution of dwarfs compared to giants in rich clusters place 
constraints on the various models of cluster evolution (such as galaxy 
harassment, Moore et al (1995)). The effect determined here is
unfortunately only suggestive and a more detailed analysis, with data
out to beyond the cluster radius, is required before any definite
conclusions can be drawn (eg. Morshidi et al. 1995).

\section{Comparison With Other Clusters}

In a previous paper (Driver {\it et al.} 1994b) a similar technique was applied
to observations of
the cluster A~963, based on data taken with the Hitchhiker parallel CCD camera
on the William Herschel Telescope at La Palma. The A~963 data had the
disadvantage of the cluster and blank fields being taken a few weeks apart and
under non-perfect conditions. Nevertheless the photometry for A~963 was shown 
to be consistent within a combined random error of 0.1 mag to the overlapping 
data of Butcher, Oemler \& Wells (1983). A~963 lies at $z =  0.206$ and is of 
richness class 3, the same as A~2554. Figure 7 shows the new A~2554 result 
overlaid with the A~963 data. The normalization has been adjusted to give equal
numbers in the well determined giant region between 
$M_{R} = -22.5$ and $-20.5$. 

Also shown on Figure 7 is data for Coma, the most comparable nearby ($z=0.024$)
cluster in terms of richness, taken from Godwin \& Peach (1977) and also 
Thompson \& Gregory (1993). The Coma data, given in $V$ and $b_{J}$ 
respectively, has been corrected to the $R_{C}$-band by the assumption of 
constant colour 
dependencies of; ($V-R_{C}$) = 0.4, and ($b_{J}-R_{C}$) = 1.7 (based on the 
tabulated colours for ellipticals as listed in Driver {\it et al.} 1994a). 
For the three clusters shown in Figure 7, we have effective distance moduli 
$D_{Coma}  = 35.8$, 
$D_{A2554}  = 39.2$ and $D_{A963} = 40.8$, including uniform k-corrections of 
$1.5 z$ (c.f. Driver {\it et al.} 1994a), the differential 
k-corrections between galaxy types being minimal for the $R$ band at low 
redshifts.

It is worthwhile digressing momentarily to discuss the other recent   
data on Coma by Bernstein {\it et al.} (1995) and Biviano {\it et al.}
(1995), these two papers both study the core region of the cluster
and find a somewhat flatter LF than that seen by Godwin \& Peach (1977)
and Thompson \& Gregory (1993). Comparing the LFs from these four published
surveys seems to show a strong environmental dependency, with the studies
of the core regions revealing typically flatter LFs than the wider field
of view surveys\footnote{The survey sizes are:
497 galaxies ($V < 17.5$, $M_{R} < -18.5$) within 1.22 sq deg for
Godwin \& Peach (1977);
1158 galaxies ($b_{J} < 20.0$, $M_{R} < -17.5$) within 3.97 sq deg for
Thompson \& Gregory (1993);
265 galaxies ($b_{J} < 18.0$, $M_{R} < -19.0$) within 0.33 sq deg for
Biviano {\it et al.} (1995); and,
$\sim 1490$ galaxies ($R < 25.5$, $M_{R} < -11.0$) within 0.015 sq deg for
Bernstein {\it et al.}. (1995).
Note in particular that only $\sim 45$ galaxies from the Bernstein CCD
sample are contained in the overlap range with our data 
($-24 < M_{R} < -16$).}
This picture is consistent with the results shown for A2554 in Figure 5,
whereby a significantly steeper density profile for the more luminous   
cluster members is seen. Clearly to derive a cluster's total LF it is 
important to survey over a significant fraction of the cluster's total
extent. Hence here we use the Thompson \& Gregory sample for our
local comparison as the sample size (over the relevant absolute magnitude
range) and field-of-view are largest.

The three data sets for the three clusters shon in Figure 7 
(spanning a range of redshift $z$ = 0.02 -- 0.2) for 
similar richness clusters can be seen to show a very strong similarity, with
a strong up turn at about $M_{R} = -19$ and a steep slope $\alpha \simeq -1.5$
faintwards of that point. (Note that $M_{R} = -19$ corresponds to about
$M_{B} = -16 + 5$ log$h$, the conventional giant/dwarf boundary). In
conjunction with other recent work on rich clusters, particularly that of
De Propris {\it et al.} (1995), this tentatively suggests that this form of 
the luminosity distribution of rich clusters may be ubiquitous.  
It is arguably best described by the sum of two 
individual luminosity functions for the separate giant and dwarf populations. 
The lines shown on Figure 7 represent a Schechter (1976) 
function for the giants 
with parameters $M^{*}_{R} = -22.3$, $\alpha = -1.0$ (cf. Efstathou et al 1988)
and a dwarf Schechter 
function with parameters 
$\phi_{*}$(Dwarfs) = $1.5 \times \phi_{*}$(Giants), $M^{*}_{R} = -18.8$, 
$\alpha = -1.7$. While we contend that a two component luminosity function
provides a more appropriate fit (see also Biviano {\it et al.} 1995;
Lobo {\it et al.} 1995), 
a single luminosity function with slope 
$\alpha \simeq -1.5$ (not shown) can also provide a reasonable description
(see Bernstein {\it et al.} 1995). 
[The reader is recommended to freely adopt either description within the bounds 
defined by Figure 7, but to beware that they give
significantly different extrapolations/implications at fainter magnitudes,
illustrating the 
peril of extrapolating any luminosity function beyond the limits of the 
available data.]

Marzke {\it et al.} (1994a,b) have recently presented evidence that the field 
galaxy LF may also show a turn up to a steeper slope at faint magnitudes
(see also Keele \& Wu 1995, and the discussion in Driver \& Phillipps 1996, though see also Ellis {\it et al.} 1996 for a contrary view).
The luminosity functions of other poorer clusters
(Ferguson \& Sandage 1991) and especially the well studied Virgo Cluster
(Sandage, Binggeli \& Tammann 1985; Binggeli, Sandage \& Tammann 1988) also present clear evidence for
both the separate population LFs and the overall steep slope at the faint end. 

\section{Evolution}

As noted, the above LFs show a remarkable similarity, with no significant 
difference in the magnitude at which the steep slope cuts in. We can thus,
very tentatively (given a sample of only 3 clusters), say that they do not show any clear evidence for evolution with cosmic epoch over the last quarter of the age of the universe (3.3 Gyr since $z = 0.2$ in our assumed 
cosmology). 
Of course, we should remember that the form of the LF may be conspiring to
hide evolution of the dwarfs. If the dwarf component happened to be less 
numerous, but brighter, in our more distant clusters, then the two effects
would cancel out in the observed LF (since we see only the power law tail of 
the dwarf luminosity distribution). Observations of further clusters of
different morphological types or densities should help to clarify this (cf.
Ferguson \& Sandage 1991; Turner {\it et al.} 1993).  

To estimate the size of effect that might be measurable, we can
shift the points for A~963 (the most distant cluster) by amounts $\Delta m$
until they become inconsistent with the line defined by the nearer clusters.
A shift (ie. assumed fading) of 0.4 magnitudes would put all the A~963 points
(in the dwarf dominated region) significantly to the faint side of the line,
so this should be readily detectable.
How much might we expect? Two possible cases can be considered.
If the light is dominated by old stellar
populations (eg. dwarf ellipticals), then a reasonable estimate
can be made via simple models of elliptical galaxy evolution, originally due to Gunn \& Tinsley (1972). In this case red giants are taken to supply most of the 
luminosity so the evolution depends on the varying number of these as stars 
progressively turn off the main sequence. For a Salpeter like stellar initial
mass function, a good approximation is $L \propto (t - t_{form})^{2/3}$ or
for our model (and early $t_{form}$) $\Delta M \simeq -2.5 log(1+z)$.
Thus by $z = 0.2$ we expect 0.2 magnitudes brightening. This level of
evolution is also seen in the much more sophisticated
passive evolution models of, eg., Bruzual \& Charlot (1993).
If, instead, the dwarf galaxies are fading irregulars, then we might expect
rather more evolution (eg. Davies \& Phillipps 1988), though this will
still be tempered by our observing at red wavelengths. 
If we set $t_{form}$ in the above equation to
correspond to, say, $z\simeq 0.5$, or use the Bruzual \& Charlot models to look
at the evolution over the first few Gyr after a starburst ends, then
we might expect $\simeq 0.^{m}6$ of evolution since $z=0.2$ (3.3 Gyr ago). 

The latter fading irregular model has previously been successfully used 
to solve the problem of the steep galaxy number counts at faint 
magnitudes and the corresponding redshift distributions (see Phillipps
\& Driver 1995 and references therein). Furthermore, evolution of the field LF 
in line with this model has been observed in deep redshift surveys (Lilly
{\it et al.} 1995; Ellis {\it et al.} 1996). It would thus be of interest 
if a larger sample of clusters continued to show no evidence for evolution
at that level, suggesting a different (recent)
evolutionary path for dwarf galaxies in cluster and field environments
(eg. Moore, Katz \& Lake 1996).

\section{Summary}

We have traced the LF of the cluster A~2554 (at $z=0.1)$ down to a 
level of $M_{R} \simeq -16$ (for $H_{0} = 50$ km s$^{-1}$ Mpc$^{-1}$), or
about $3 \times 10^{-3} L_{*}$. The LF shows the characteristic upturn at 
$M_{R} \simeq -19$ seen in previous cluster observations. The overall LF
can be well fitted by two (Schechter type) components, the giants having the
conventional $\alpha \simeq -1$, the dwarfs a steeper $\alpha \simeq -1.7$. 
An alternative fit with a single $\alpha \simeq -1.5$ is also possible
within the errors.

The comparison with the LFs for Coma ($z = 0.02$) and A~963 ($z = 0.2$)
shows a remarkable similarity, suggesting (a) a ubiquitous dwarf population
in rich clusters and (b) no evidence for significant
evolution of the cluster dwarfs (at least at red wavelengths)
over the last 3.3 Gyr. The level of evolution expected if cluster dIs
have evolved to dEs since $z \simeq 0.5$, as required in some successful models of the evolution
of their counterparts in the field, should be detectable in this type of data,
though passive evolution probably would not.

In subsequent papers we will consider a further large sample ($\sim 15$)
rich clusters, observed to similar depths to the A~2554 data presented here.
With the full set of clusters we will be able to investigate whether
the steep slope cuts in at the same absolute magnitude in each case. If it does
not, it will be of great interest to see whether the point where dwarf
domination starts changes systematically with $z$. 
Of course, as pointed out earlier, if the LF
really is of two component form, then the point where the steepening takes
place depends on both the characteristic luminosity for the dwarfs (which might
evolve) $and$ on the ratio of dwarfs to giants in the cluster (which
might change with cluster properties). In particular it will be important to
see whether the Bautz-Morgan type (effectively an indicator of cluster density
as opposed to richness) influences the dwarf to giant ratio {\it
at given $z$} and
hence the overall shape of the LF. 

\section*{Acknowledgments}

We thank Warrick Couch for his input to the cluster photometry programme and
for comments on the current paper. We also
thank Paul Bristow, John Bryn Jones, Igor Karachentsev,
Bahram Mobasher, Jean Marc Schwartzenberg and Ray Sharples for
useful discussions. Some of this work was carried out while SPD was at the
Department of Physics and Astronomy, Arizona State University. SP thanks the
Royal Society for support via a University Research Fellowship. SPD
acknowledges the Australian Research Council for support.

\section*{References}

Abell G., 1958, ApJS, 3, 1.\\
Abell G., Corwin H.G., Olowin R., 1989, ApJS, 301, 83.\\
Barger A.J., Aragon-Salamanca A., Ellis R.S.,  Couch W.J., Smail I., Sharples R.M., 1996, MNRAS, 279, 1\\
Bernstein G.M., Nichol R.C., Tyson J.A., Ulmer M.P., Wittman D., 1995
AJ, 110, 1507\\
Binggeli B., Sandage A., Tammann G.A., 1988, ARA\&A, 26, 509\\
Biviano A., Durret F., Gerbal D., Le Fevre O., Lobo C., Mazure A., Slezak E., 
1995, A\&A, 297, 610\\
Bristow P.D., 1996, Ph.D. Thesis, University of Wales, Cardiff.\\
Broadhurst T.J., Ellis R.S., Shanks T., 1988, MNRAS, 235, 827\\
Bruzual G.A., Charlot S., 1993, ApJ, 405, 538\\
Butcher H., Oemler A., 1984, ApJ, 376, 404\\
Butcher H., Oemler A., Wells D.C., 1983, ApJS, 52, 183\\
Colless M.M., 1995, in  Maddox S.J., Aragon-Salamanca A. eds, Wide Field Spectroscopy and the Distant Universe (World Scientific), p~263\\
Couch W.J., Jurecevic J.S., Boyle B.J., 1993, MNRAS, 260, 241\\
Couch W.J., Sharples R.J., 1987, MNRAS, 229, 423\\
Davies J.I., Disney M.J., Phillipps S., Boyle B.J., Couch W.J., 1994, 
MNRAS, 269, 349\\
Davies J.I., Phillipps S., 1988, MNRAS, 233, 553\\
Driver S.P., Phillipps S., Davies J.I., Morgan I., Disney M.J., 1994a, MNRAS,
266, 155.\\
Driver S.P., Phillipps S., Davies J.I., Morgan I., Disney M.J., 1994b, MNRAS,
268, 393.\\
Driver S.P., Phillipps, S., 1996, ApJ, 469, 529 \\
De Propis, R., Pritchet, C.J., Harris, W.E., McClure, R.D., 1995, ApJ, in press\\
Efstathiou G., Ellis R.S., Peterson B.A., 1988, MNRAS, 232, 431.\\
Ellis R.S., Colless M., Broadhurst T.J., Heyl J., Glazebrook K., 1996, 
MNRAS, 280, 235 \\
Ferguson H., Binggeli B., 1995, A\&AR, 6,67\\
Ferguson H., Sandage A., 1991, AJ, 101, 765. \\
Godwin J., Peach J.V., 1977, MNRAS, 181, 323.\\
Godwin J., Metcalfe, N., \& Peach, J.V., 1983, MNRAS, 202, 113 \\
Gunn J.E., Tinsley B.M., 1972, ApJ, 203, 52\\
Keele W.C., Wu, W. 1995, AJ., 110,129\\
Landolt A.R., 1992, AJ, 104,340\\
Leir A.A.,  van den Bergh,S., 1977, ApJS, 34, 38\\
Lilly S.J., Tresse L., Hammer F., Crampton D., Le Fevre O., 1995, ApJ,
455, 108\\
Lobo C., Biviano A., Durret F., Gerbal D., Le Fevre O., Mazure A., Slezak E., 
1995, in 
Giuricin G., Mardirossian F., Mezzetti M. eds, Observational Cosmology: 
From Galaxies to Galaxy Systems\\
Marzke R.O., Huchra J.P., Geller M.J., 1994a, ApJ, 428, 43\\
Marzke R.O., Geller M.J., Huchra J.P., Corwin H.G., 1994b, AJ, 108, 437\\
McGaugh S.S., 1994, Nature, 367, 538.\\
Metcalfe N., Shanks T., Fong R., Roche N., 1995, MNRAS, 273, 257\\
Moore B., Katz N., Lake G., 1995, in Bender R., Davies R.L. eds, New Light on Galaxy Evolution, (Kluwer, Dordrecht), p~203\\
Morshidi, Z., Smith, R.M. and Davies, J.I., 1995, in Bender R., Davies R.L. eds, New Light on Galaxy Evolution, (Kluwer, Dordrecht), p~423\\
Phillipps S., Driver S.P., 1995, MNRAS, 274, 832.\\
Phillipps S., Driver S.P., Smith R.M., 1995a, in Giuricin G., Mardirrosian F., Mezzetti M. eds, 
Observational Cosmology:
From Galaxies to Galaxy Systems (SISSA, Trieste)\\
Phillipps S., Driver S.P., Smith R.M., 1995b, Ap. Lett. \& Comm., in press\\
Sandage A., Binggeli B., Tammann G.A., 1985, AJ, 90, 1759\\
Schade D., Ferguson H.C., 1994, MNRAS, 267, 889\\
Schechter P., 1976, ApJ, 203, 297\\
Schwartzenberg J.M., Phillipps, S., Smith R.M., Couch W.J., Boyle B.J., 1995,
MNRAS, 275, 121\\
Schwartzenberg J.M., 1996, Ph.D. thesis, Univ. of Bristol \\
Thompson L.A., Gregory S.A., 1993, AJ, 106, 2197 \\
Turner J.A., Phillipps S., Davies J.I., Disney M.J., 1993, MNRAS, 261, 39

\pagebreak

\section*{Figures}

{\bf Figure 1.} Grey scale representation of the AAT field around
Abell 2554. The $17'$ field of view here corresponds to about 3 Mpc.
\\
\\
{\bf Figure 2.} Isophotal area versus isophotal magnitude plot for all detected
images in the A~2554 frame. The selection limits of 4 pixels (dashed
line) and $7.5\sigma$ (solid line) are also shown, together with the loci of 
galaxies with scale size of 2arcsec (dotted line).
\\
\\
{\bf Figure 3.} The correction required for the diminishing area available to 
faint objects because of the area occupied by brighter objects. Note that the
cluster correction (solid squares) is significantly greater than the 
field correction (solid triangles). The lines show the correction for the core
(short dashed) and outer regions (long dashed) of the cluster sight-line 
demonstrating that the distribution of cluster galaxies is concentrated towards
the field centre.
\\
\\
{\bf Figure 4.} (upper panel) R-band number-counts for A~2554 (solid squares)
and SGP (open squares). Also shown (solid line) is our best fit line to the
data of Metcalfe {\it et al.} (1995) corrected to $R_{c}$.
(lower panel) The derived luminosity function for A~2554 by subtraction of
the SGP data (filled squares) and by the Metcalfe et al data (filled 
triangles).
\\
\\
{\bf Figure 5.} Integrated radial density profile of A~2554. The sample has 
been divided into two to represent the distribution of giants (circles) and 
dwarfs (triangles). The mean field density first subtracted and the data 
are scaled such that the total excess for giants or dwarfs is unity.
\\
\\
{\bf Figure 6.} The isodensity structure of A~2554 for the giants
(a) and the dwarfs (b) and for the blank control SGP field bright (c)
and faint galaxies (d). The gray scale is set to display the overdense regions 
and the same scaling is used for all plots. The contours are as listed in the 
text. Fig. 6c mimics Fig. 6a showing that the
dwarf galaxies are distributed like the giants but may have a flatter
radial distribution. Figs 6b and 6d show that the structure
seen for giants and dwarfs in A~2554 is significant when compared to the 
typical structure seen in a blank sight-line.
\\
\\
{\bf Figure 7.} Comparison of the A~2554, A~963 and A~1367 (Coma) Cluster LFs.
All three clusters show a remarkable similarity in their galaxy luminosity 
distributions even though they span a wide range in redshift (0.02 -- 0.2).
The dotted line shows a Schechter LF with parameters  $M^* = -22.5$ and
$\alpha = -1.0$ whilst the dashed line is a Schechter LF with
$M^*(R) = -19.5$ and $\alpha = -1.8$ and a $\phi^*$ of twice that of the dashed
line.

\pagebreak

\begin{figure}[p]
\vspace{10.0cm}
{\bf FIGURE 1 not included}
\vspace{10.0cm}
\end{figure}

\begin{figure}[p]
\centerline{\hspace{0.0cm} \psfig{file=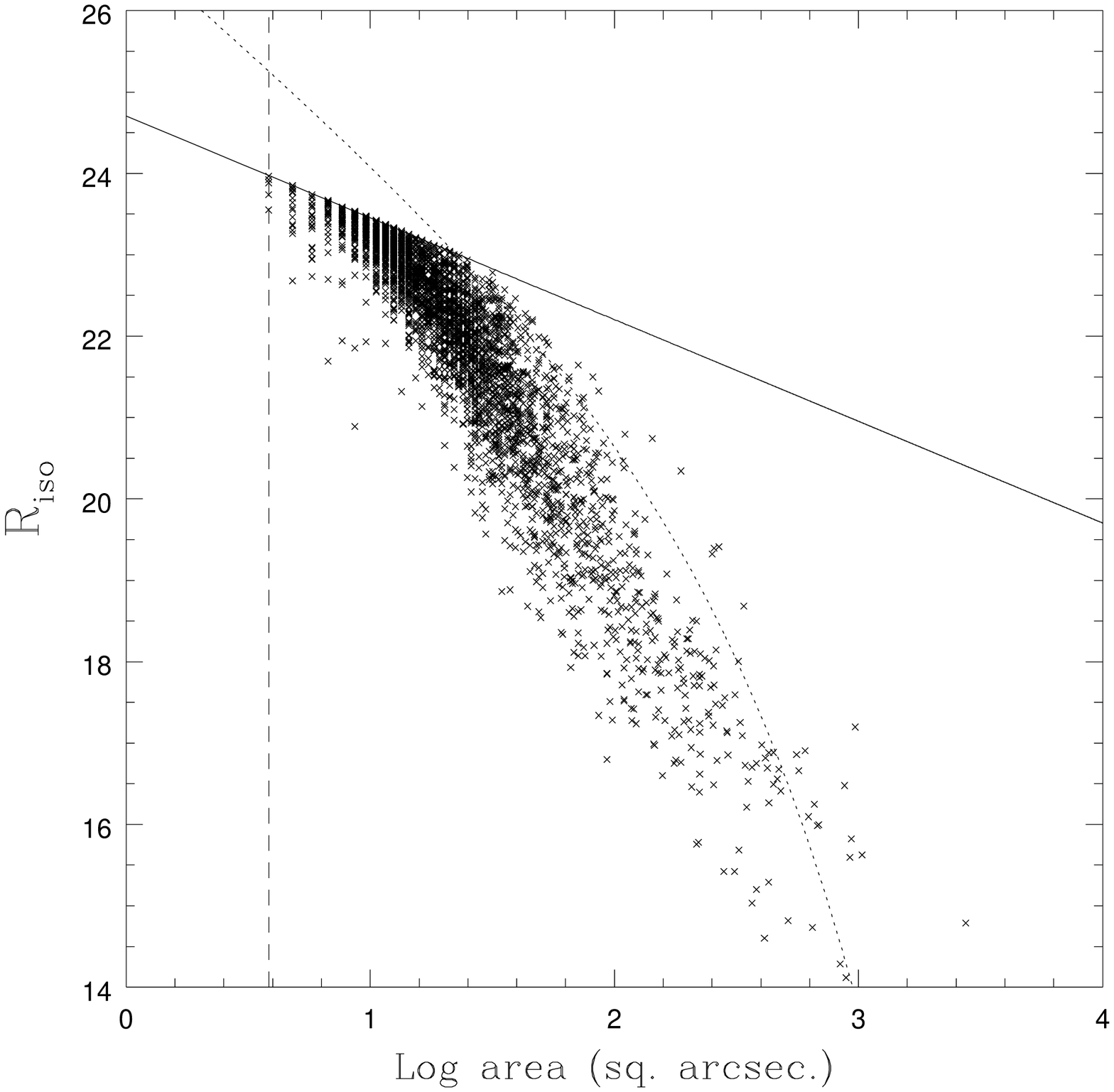,height=200mm,width=180mm}}
\end{figure}

\begin{figure}[p]
\centerline{\hspace{0.0cm} \psfig{file=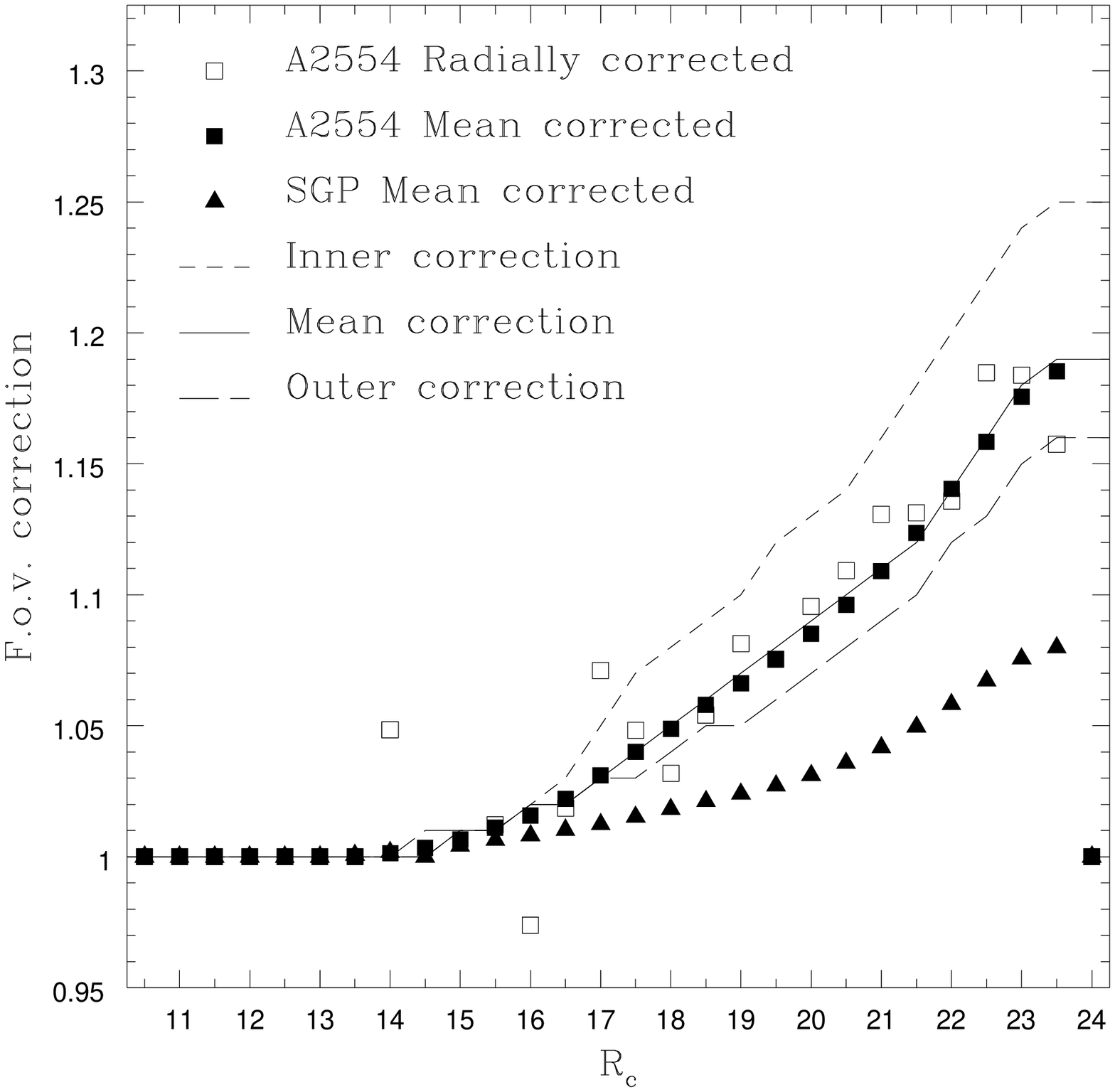,height=200mm,width=180mm}}
\end{figure}

\begin{figure}[p]
\centerline{\hspace{0.0cm} \psfig{file=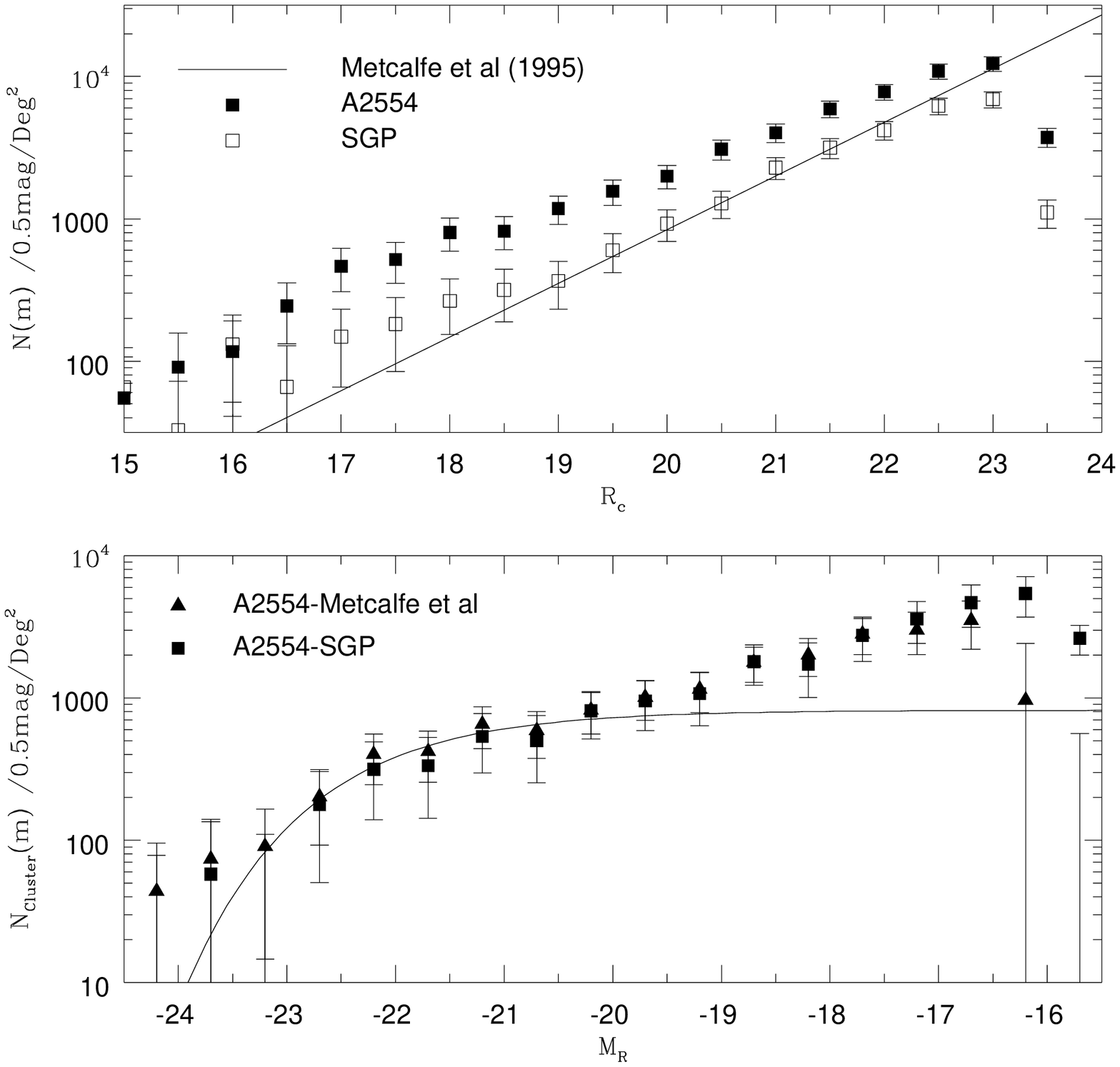,height=200mm,width=180mm}}
\end{figure}

\begin{figure}[p]
\centerline{\hspace{0.0cm} \psfig{file=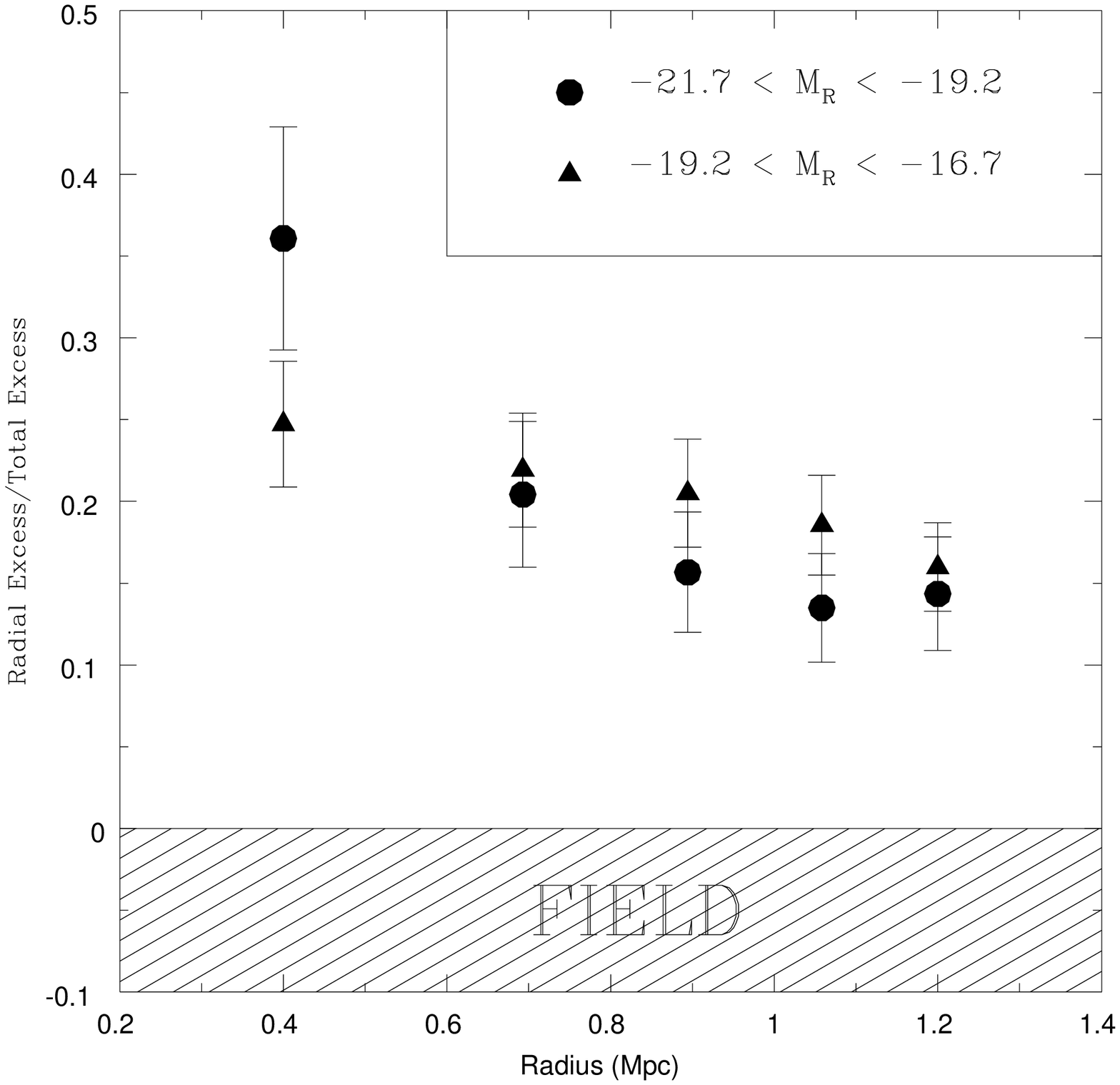,height=200mm,width=180mm}}
\end{figure}

\begin{figure}[p]
\vspace{-3.0cm}
\centerline{\hspace{0.0cm} \psfig{file=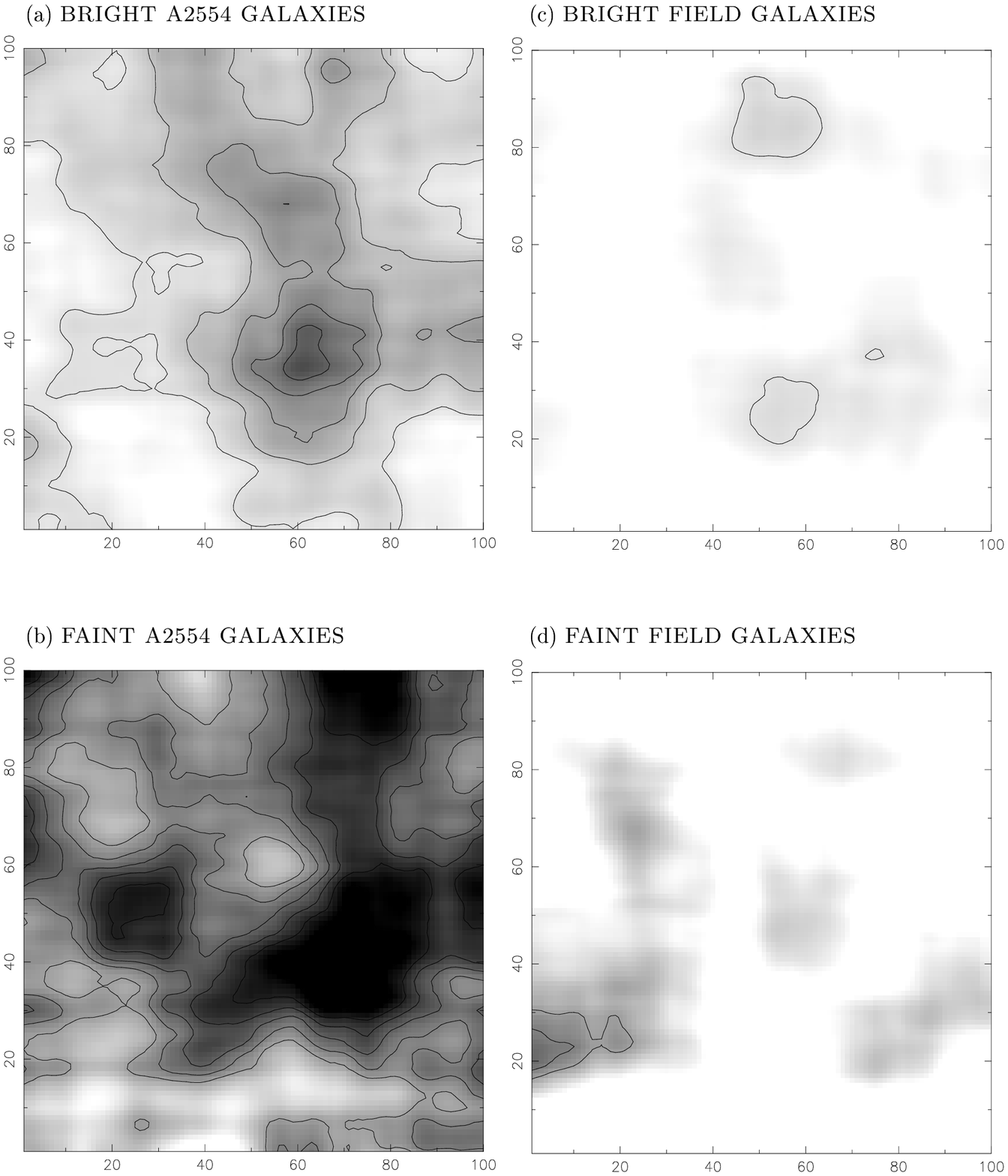,height=260mm,width=210mm}}
\end{figure}

\begin{figure}[p]
\centerline{\hspace{0.0cm} \psfig{file=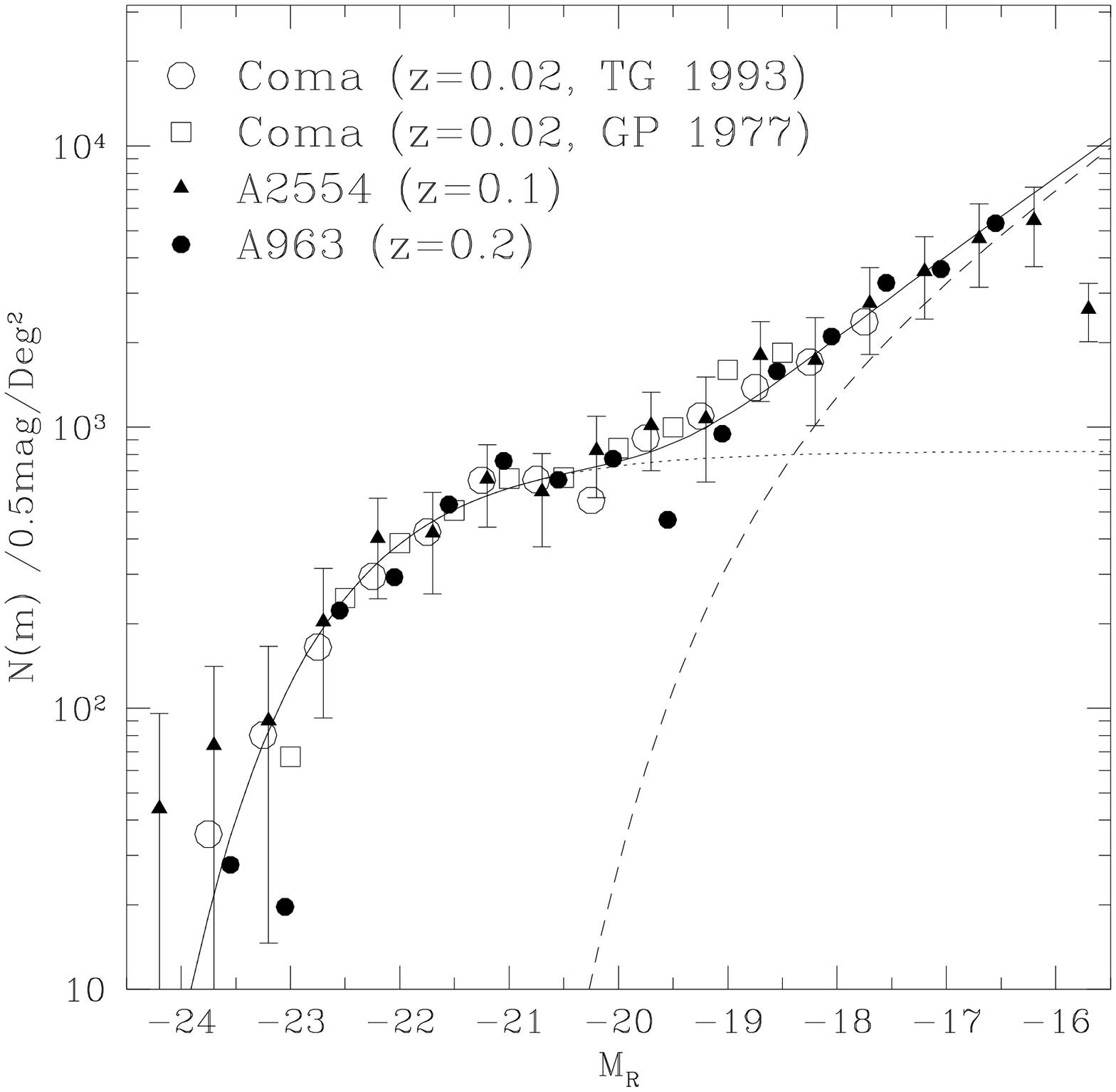,height=200mm,width=180mm}}
\end{figure}

\end{document}